# Screening for REM Sleep Behaviour Disorder with Minimal Sensors

Navin Cooray[1], Fernando Andreotti[1], Christine Lo[2], Mkael Symmonds[3], Michele T.M. Hu[2], & Maarten De Vos[1]

[1]University of Oxford, Institute of Biomedical Engineering, Dept. Engineering Sciences, Oxford, UK.

[2]Nuffield Department of Clinical Neurosciences, Oxford Parkinson's Disease Centre (OPDC), University of Oxford, UK.

[3]Department of Clinical Neurophysiology, Oxford University Hospitals, John Radcliffe Hospital, University of Oxford, UK.

Author to whom all correspondence should be directed is NC (navin.cooray@eng.ox.ac.uk)

## Abstract

Objective: Rapid-Eye-Movement (REM) sleep behaviour disorder (RBD) is an early predictor of Parkinson's disease, dementia with Lewy bodies, and multiple system atrophy. This study investigates a minimal set of sensors to achieve effective screening for RBD in the population, integrating automated sleep staging (three state) followed by RBD detection without the need for cumbersome electroencephalogram (EEG) sensors.

Methods: Polysomnography signals from 50 participants with RBD and 50 age-matched healthy controls were used to evaluate this study. Three stage sleep classification was achieved using a Random Forest (RF) classifier and features derived from a combination of cost-effective and easy to use sensors, namely electrocardiogram (ECG), electrooculogram (EOG), and electromyogram (EMG) channels. Subsequently, RBD detection was achieved using established and new metrics derived from ECG and EMG metrics.

Results: The EOG and EMG combination provided the best minimalist fully automated performance, achieving 0.57±0.19 kappa (3 stage) for sleep staging and an RBD detection accuracy of 0.90±0.11, (sensitivity, and specificity 0.88±0.13, and 0.92±0.098). A single ECG sensor allowed three state sleep staging with 0.28±0.06 kappa and RBD detection accuracy of 0.62±0.10.

Conclusions: This study demonstrated the feasibility of using signals from a single EOG and EMG sensor to detect RBD using fully-automated techniques.

Significance: This study proposes a cost-effective, practical, and simple RBD identification support tool using only two sensors (EMG and EOG), ideal for screening purposes.

## Highlights

- This study proves the feasibility of an effective screening tool that simply uses an EMG, EOG or ECG sensors for fully automated RBD detection.
- The REM sleep stage was accurate and detected reliably in individuals with RBD without the use of EEG.
- Automated sleep staging achieved high positive predictive value of REM sleep stage and ensured RBD detection performance was also high.

This research was supported by the Research Council UK (RCUK) Digital Economy Programme (Oxford Centre for Doctoral Training in Healthcare Innovation -- grant EP/G036861/1), Sleep, Circadian Rhythms & Neuroscience Institute (SCNi -- 098461/Z/12/Z), Rotary Foundation, National Institute for Health Research (NIHR) Oxford Biomedical Research Centre (BRC), Engineering and Physical Sciences Research Council (EPSRC -- grant EP/N024966/1), and Parkinson's UK: Targeting the pathological pathways to Parkinson's (grant J-1403). The content of this article is solely the responsibility of the authors and does not necessarily represent the official views of the RCUK, SCNi, NIHR, BRC or the Rotary Foundation.

## Conflict of interest

None of the authors have potential conflicts of interest to be disclosed.

## 1.  Introduction

In the search for an early predictor for the onset of Parkinson's disease, few are as promising as Rapid-Eye-Movement (REM) Sleep behaviour disorder (RBD). Recent literature estimates 73.5% conversion of RBD to an overt neurodegenerative syndrome after a 12-year follow-up with a 6% risk of conversion per year (Postuma et al. 2019). To further explore the mechanisms behind these conversions and to develop neuroprotective therapies, larger cohorts of individuals with RBD must be identified. The greatest hurdle to designing such studies is the slow and arduous RBD diagnostic process involving polysomnography (PSG) recordings within sleep clinics, which are often under-staffed and oversubscribed (severely limiting availability). PSG recordings are expensive, time consuming to setup, can be uncomfortable and rely on labour intensive visual inspection of PSG data. Previous studies have demonstrated that automated sleep staging and RBD detection was possible from a limited PSG consisting of a electroencephalogram (EEG), electrooculogram (EOG), and electromyogram (EMG) sensor (Cooray et al. 2018, 2019). This limited framework validates the possibility of a practical and comfortable take-home device for sleep evaluation, of which only a limited number of clinical grade devices supported by literature exist (Mikkelsen et al. 2017; Sterr et al. 2018).

Recent sleep research is uncovering an interplay between the central nervous system (CNS) and the autonomic nervous system (ANS) that is regulated throughout sleep and its various stages (de Zambotti et al. 2018). The ANS is responsible for regulating the majority of the body's internal systems, such as blood pressure, breathing, body temperature, digestion, urine, and myocardial function. Of particular interest is the change in heart rate variability (HRV) during sleep and how it might be interpreted for automated three state sleep staging (Redmond and Mcnicholas 2007; Mendez et al. 2010; Xiao et al. 2013; Ebrahimi et al. 2015; Fonseca et al. 2015; Malik et al. 2018; Yücelbaş et al. 2018). For these studies automated sleep staging from ECG based features were achieved for three sleep states, wake (W), REM and non-REM (REM). Additionally, recent literature indicated cardiac autonomic dysfunction as a characteristic of RBD (Postuma et al. 2010; Sorensen et al. 2013; Bugalho et al. 2018). Postuma et al. (2010) demonstrated through RR standard deviation and low frequency components that there is a clear autonomic dysfunction in RBD participants compared to healthy controls. Similarly, Sorenson et al. (2013) illustrated attenuated sympathetic activity in heart rate variability in participants with PD and/or RBD compared to healthy controls. Bugalho et al. (2018) attributed to RBD participants a reduction in parasympathetic modulation of heart rate variability in relation to sleep stage transitions. Heart rate and heart rate variability (HRV) are commonly used to evaluate components of ANS functionality (Andreotti et al. 2018), conventionally measured by electrocardiogram (ECG) sensors. ECG sensors are ubiquitous and user-friendly and therefore provide an opportunity to further economise sleep staging and RBD detection by excluding the need for cumbersome and expensive EEG sensors.

Furthermore, single sensor studies in sleep staging have been proposed for EOG and EMG sensors, which aim to be more economical whilst producing similar performances to EEG electrodes (Virkkala et al. 2007, 2008; Yetton et al. 2016). With respect to RBD detection, our previous work utilised established objective techniques to evaluate REM sleep without atonia (RWSA) and demonstrated how they could be improved using a combination of EMG metrics that incorporate sleep architecture (Cooray et al. 2018, 2019). We will investigate the inclusion of ECG based metrics that exploit changes in ANS, as a substitute or enhancement for RBD detection. Consequently, using a simplified combination of ECG, EOG and EMG sensors, this study explored the feasibility of simpler RBD screening, being minimalistic while proving as effective at identifying RBD compared to conventional full PSG techniques.

## 2. Data

PSG recordings used in this study were collected from several sources and included participants diagnosed with RBD and age-matched healthy controls (HCs), detailed in Table 1. The Montreal Archive of Sleep Studies (MASS) cohort one (O'Reilly et al. 2014) provided 53 HC individuals, where three were excluded due to their poor ECG signal fidelity (specifically, participants labelled SS01, SS10, and SS18). Clinically diagnosed participants with idiopathic RBD were collated from the Physionet Cyclic Alternating Pattern (CAP) sleep database (22 RBD participants) (Goldberger et al. 2000; Terzano et al. 2001) and a private database of RBD participants from the John Radcliffe (JR) hospital (35 participants), Nuffield Department of Clinical Neurosciences at the University of Oxford. While the JR dataset included two nights of full PSG recordings for each participant, this study simply used the second night, where available (only two participants had a single night recording). Eight participants from the CAP database were excluded because one was a duplicate recording (RBD11) and seven had secondary RBD, including five diagnosed with PD (RBD1, RBD2, RBD3, RBD7, and RBD9), one with multiple systems atrophy (RBD13), and one with Lewy body disease (RBD5). Consequently, this combined RBD cohort provided a total of 50 RBD participants, of which two were taking Clonazepam to treat their condition, while ten participants were taking anti-depressants (Citalopram, Venlafaxine, Sertraline, Duloxetine, and Lorazepam) to treat REM sleep without atonia (RSWA). All RBD participants had an apnoea-hypopnea index (AHI) of less than 7.1 and were evaluated as being mild and unremarkable, except for one participant (RBD1) from the CAP sleep database with a severe index of 30/h (previously excluded for secondary RBD). Five RBD participants with concurrent RBD and obstructive sleep apnoea (OSA) wore a continuous positive airway pressure ventilator at the time of their PSG recordings, which would benefit their AHI score. This study complied with the requirements of the Department of Health Research Governance Framework for Health and Social Care 2005 and was approved by the Oxford University hospitals NHS Trust (HH/RA/PID 11957). These two RBD datasets were combined to balance the HC recordings provided by the MASS database. Furthermore, this combination provides an opportunity to evaluate how generalizable this study is over differing datasets annotated at different institutions. Once more by using openly available datasets (MASS and CAP), this study can be reproduced in combination with the toolbox provided at https://github.com/navsnav/Minimal-RBD-Sleep-Detection.

Annotations of the PSG recordings were completed by experts using either the Rechtschaffen and Kales (R&K) (Rechtschaffen and Kales 1968) or AASM guidelines (Iber et al. 2007). For the purposes of this study, these set of rules are easily converted to three stages (REM, NREM, and W), where NREM is defined by S1, S2, S3, S4, N1, N2 or N3. To explore the most economical combination of sensors, the following signals were analysed:

- 1 ECG (2 electrodes - Einthoven derivation)
- 1 EOG (2 electrodes – bipolar signal)
- 1 EMG (2 electrodes – submentalis)

Numerous studies have used limited sensors to assess automated sleep staging, but for this study we applied automated sleep staging on healthy and RBD participants followed by automated RBD identification.

## 3. Method

### 3.1 Pre-processing

To reduce the impact of noise and artefacts, PSG signals were pre-processed. Firstly, all ECG, EOG and EMG signals were resampled at 200Hz. The EOG signal was pre-processed with a $500^{th}$ order band pass finite impulse response (FIR) filter with a cut-off frequency of 0.3Hz and 40Hz. The EMG signal was filtered with a $500^{th}$ order notch filter at 50Hz and 60Hz (because the recordings were sourced from either Europe or Canada), in addition to a $500^{th}$ order band pass FIR filter between 10Hz and 100Hz. The ECG signal was filtered by a $10^{th}$ order Butterworth band pass filter between 5Hz and 45Hz.

### 3.2 Feature Extraction

Feature extraction for EOG and EMG signals were as described in our previous study (Cooray et al. 2019), based on established literature in automated sleep staging (Güneş et al. 2010; Koley and Dey 2012; Liang et al. 2012; Lajnef et al. 2015; Yetton et al. 2016). Literature that described ECG features for automated sleep staging provided motivation for this study (Redmond and Mcnicholas 2007; Mendez et al. 2010; Xiao et al. 2013; Ebrahimi et al. 2015; Fonseca et al. 2015; Yoon et al. 2017; Malik et al. 2018; Yücelbaş et al. 2018) and are summarised in Table 2. For this study the pre-processed ECG signal was segmented into five-minute epochs, often used for HRV analysis (Ebrahimi et al. 2015). Features were then derived after Pan-Tompkins QRS detection (Pan and Tompkins 1985) from each 30-second segment. These features were then averaged across a moving 150-second sliding window, demonstrated in other similar studies (Redmond and Mcnicholas 2007; Xiao et al. 2013; Fonseca et al. 2015).

### 3.3 Automated Sleep Stage Classification

The Random Forest (RF) algorithm (Breiman 2001) has proven effective for automated sleep staging and was used for this study. The classifier was trained to classify 30-second epochs into one of three sleep stages (REM, NREM and W) using a combination of 25 EOG features, 17 EMG features and 75 ECG features, as described in Table 2 (the number of trees was set to 500, $m_{try} = \sqrt{M}$ (rounded down) randomly selected features, where $M$ is the total number of features (detailed in Table 4)). For each of the three sleep stages the classifiers were evaluated using macro-averaged sleep stage accuracy, sensitivity, specificity, and Cohen-Kappa score (three stage) by using 10-subject-fold cross-validation with an even split between healthy and RBD participants. Additionally, multi-stage classification was assessed by Cohen's Kappa (Cohen 1960), a metric often used in sleep staging to evaluate the agreement between manual and automated annotations. These performances were compared to results obtained using EEG, EOG and EMG features described in Cooray et al (2019). Our previous work demonstrated that accurate automated sleep staging (especially REM detection) yielded a high performance in RBD detection.

### 3.4 RBD Detection

In RBD diagnosis, literature stipulates that after the identification of REM sleep, RSWA must be confirmed visually (Sateia 2014). In our previous work we demonstrated that established features, which quantify RSWA in combination with sleep architecture, provided RBD detection performances approaching that of the gold standard (manual clinical diagnosis) by using RF classifiers (Cooray et al. 2018, 2019). This study emulated previous results and compared them to RBD detection using additional ECG based metrics in isolation and combination with EMG based metrics through RF classification.

RF classifiers were used to achieve RBD detection, as per the combinations detailed in Table 5. Classifiers were trained using 500 trees; $m_{try} = \sqrt{M}$ (where M is the number of features detailed in Table 5), and evaluated using a 10-subject-fold cross-validation scheme. In addition to the features detailed in our previous study, we proposed new ECG features that capture changes in heart rate variability between sleep stages (see Table 3**Error! Reference source not found.**).

A summary of all ECG based metrics that incorporate sleep architecture and HRV in order to identify RBD are detailed in Table 3. Specifically, to quantify the variability, the mean sample entropy for each REM epoch was calculated for every participant. Similarly the standard deviation of the RR intervals for every REM epoch was averaged for all participants. Often low frequency and high frequency components of RR intervals are used to describe sympathetic and parasympathetic activity in the autonomic nervous system, and as a result the mean peak low frequency and high frequency was calculated for all participants. To capture the changes in heart rate variation between sleep stages, the ratio between REM and NREM values was calculated for the mean RR interval. The low frequency and high frequency ratio is a popular feature used in heart rate variability, and the ratio between NREM and REM epochs was calculated for each participant as an RBD metric. Literature has also suggested that PD is significantly comorbid with atrial fibrillation (AF) (Hong et al. 2019), consequently the metrics termed irregular index and origin count (Sarkar et al. 2008; Oster and Clifford 2015) were used to quantify irregularly irregular heartbeats during NREM and REM sleep. These ECG based metrics were used to automatically identify RBD participants and were compared to metrics from our previous studies (Cooray et al. 2018, 2019).

Furthermore the RBD detection performance for these metrics were evaluated using manually and automatically annotated sleep staging using the best combination of minimal sensors (see Table 4), in order to validate their use in an end-to-end RBD screening support tool. The impact of automated sleep staging on the calculation of these metrics was also carefully analysed.

The fidelity and consistency of these ECG based metrics derived though manually and automatically annotated sleep staging were evaluated by Bland and Altman (B&A) plots (Bland and Altman 1986, 1995, 1999). B&A plots were introduced to describe the agreement between two quantitative measures as an improvement to simple correlation factors. While correlation factors provide the strength of a linear relationship, they do not imply a good agreement between two methods. The B&A plot measures agreement by plotting the mean against the difference of both quantitative methods (Bland and Altman 1986), while constructing the limits of agreement (where Bland and Altman et al. recommend that 95% of all data points fall within two standard deviations of the mean difference). Furthermore, B&A plots detail the bias between the mean differences of both quantitative methods and their significance. For this study B&A plots were used to establish a level of agreement between RBD metrics, derived using manually and automatically annotated sleep stages within a certain agreement interval. The decision on whether these intervals are acceptable must be decided by clinical evaluation and in our case through the performance of automated RBD detection with metrics derived from automated sleep staging.

## 4. Results

The following results are presented in three sections, detailing 1) automated sleep stage classification, 2) RBD detection using automatically annotated sleep stages, and 3) Bland and Altman plots for RBD metrics using automatically/manually annotated sleep stages.

### 4.1 Automated Sleep Stage Classification

The best performance of automated sleep staging was provided by all three signals (EOG, ECG, and EMG) attaining an agreement score of 0.58 (0.70 and 0.48 for the individual HC and RBD cohort, respectively). These results are also comparable to our previous study that included EEG features (Cooray et al. 2019), shown in Figure 1 as combination Z3 (EEG, EOG, and EMG). Once again it is clear that the automated sleep staging performs considerably better on HC participants than RBD participants, echoing the results from our previous study (Cooray et al. 2019). The best performing combination of two sensors proved to be the EOG and EMG sensors (C2), followed by the EOG and ECG combination (A2) achieving a 0.57 and 0.51 agreement score, respectively. While combining ECG and EMG features attained an agreement score of 0.41. The best single sensor performance was given by EOG features (see Figure 1 B1), achieving a three-stage score of 0.50, considered moderate agreement (Landis and Koch 1977). The single ECG and EMG sensor (A1 and C1, respectively) had limited success, never exceeding fair agreement (0.30 and 0.28, respectively).

The ranked feature importance in REM detection are detailed in Figure 2, for (a) ECG features and for (b) combined EOG, ECG and EMG features. From Figure 2 (a) it appears elapsed time is the most important, followed by zero crossing interval (ZCI) and RR interval based features. In Figure 2 (b) EOG features dominate the list of important features, while EMG and ECG features make an appearance in the top 30.

### 4.2 RBD Detection with ECG Metrics

The performances of RBD detection by ECG and EMG metrics are depicted in Figure 3. ECG metrics proved they are able to distinguish RBD participants with an accuracy, sensitivity, and specificity of 0.62±0.20, 0.56±0.23, and 0.68±0.24, respectively. The combination of EMG metrics and sleep architecture (detailed in our previous study) achieved an accuracy, sensitivity and specificity of 0.93±0.09, 0.92±0.13, and 0.94±0.092, respectively. Combining EMG and ECG metrics for RBD detection provided a similar accuracy, sensitivity, and specificity of 0.93±0.10, 0.94±0.092, and 0.92±0.13, respectively.

The ranking of ECG and/or EMG metrics for RBD detection are depicted in Figure 4. Simply using ECG metrics (Figure 4 (a)) indicated that NREM origin count (a measure or regular heart beats) and NREM irregular index (IrrIndex, a measure of irregularly irregular heartbeats) and the ratio or NREM sleep were the most valuable. Figure 4 (b) ranks EMG metrics, where the atonia index ratio (between NREM and REM), atonia index (REM) and the fractal exponent ratio (between NREM and REM) proved to be the most important. ECG and EMG metrics used in combination (Figure 4 (c)), demonstrated that EMG metrics remain the best identifiers in a similar order to Figure 4 (b).

Figure 5 details the best RBD detection performance from automated sleep staging classifiers using a combination of EOG, ECG and EMG features. The best performance (F1 score) was given by C2D2 (three sensors - EOG and EMG for sleep staging followed by EMG and ECG for RBD detection), followed by C2E1 combination (two sensors - EOG and EMG for sleep staging followed by EMG for RBD detection) with an accuracy, sensitivity, specificity and F1 score of 0.90±0.11, 0.88±0.13, 0.92±0.098, and 0.90±0.12. These results are similar and only marginally less than when using a cumbersome EEG combination (Cooray et al. 2019), depicted by Z3 (Figure 5).

### 4.3 RBD Metrics from Automated and Annotated Sleep Staging (Bland-Altman)

The impact of RBD metrics calculated from automated sleep staging from a single EOG and EMG is evaluated by the Bland-Altman plots given in Figure 6. A selection of important RBD metrics (provided by Figure 4) appear to provide agreeable results where the limits of agreement are represented by horizontal dotted lines. The mean difference bias is defined by the solid black line in Figure 6, where of the top three EMG and ECG based metrics all appear to have a slight bias when using automated sleep staging (shaded area around mean difference includes zero).

## 5. Discussion

The aim of this study was to investigate a minimalistic approach for identifying RBD through a fully automated pipeline. This study excluded the use of cumbersome-to-apply EEG sensors and instead focused on a combination of ECG, EOG and EMG sensors to achieve automated three-stage sleep classification and RBD identification. The use of EOG and EMG sensors for automated sleep staging proved the most effective, achieving REM detection that was comparable with the literature and near human expert annotation (Figure 1). Not surprising given visual staging (especially REM) relies heavily on EOG and EMG signals. Furthermore this study validated the use of ECG based RBD metrics to identify RBD using an RF classifier (Figure 3). Once more the performance of RBD detection remained high when using automated sleep staging from EOG and EMG features (Figure 5 and Table 7). This can be attributed to the high positive prediction value and specificity provided by the automated sleep staging that facilitated congruent RBD metrics (Figure 6).

As has been established in literature, the inter-rater variability of manual sleep staging is high and is exacerbated further by sleep disorders that provide additional variation in sleep characteristics (Danker-Hopfe et al. 2004, 2009). This was also demonstrated in our previous study, where automated sleep staging performed better on HCs than RBD participants (Cooray et al. 2019) and was again observed in this study (for all combination of sensors). For a single ECG sensor, sleep staging was particularly difficult for RBD participants.

The automated three-stage classification performance of ECG features (Table 6) on a combined dataset (kappa of 0.30), reflected the lower range of results achieved by similar studies (kappa between 0.35 and 0.73) (Redmond and Mcnicholas 2007; Xiao et al. 2013; Fonseca et al. 2015; Yücelbaş et al. 2018). However these studies benefit from smaller datasets focusing on relatively younger HCs (participants of 28 to 48 people with a mean age that varies from 41 to 43). Yücelbas et al. (2018) achieved the highest score of 0.74, but even this technique applied to another dataset achieved a kappa of 0.43 and for participants with OSA scored between 0.52 and 0.57, indicating a high degree of variability (Yücelbaş et al. 2018). Xiao et al. (2013) had the added benefit of using manually annoated ECG segments, termed "stationary", achieving a score of 0.47 (Xiao et al. 2013).

Interestingly, while this study acheives a kappa score of 0.30 (ECG features) on a mixed cohort of 100 participants, the RBD cohort severely underperforms with respect to REM detection. This was due to a misclassification of REM states for NREM and wake, potentially caused by changes to HRV regulation and from movement artefacts, which might be attributed to symptoms of RBD. The feature importance for REM detection, detailed in Figure 2 (a), ranks elapsed-time as more indicative than ECG features. Elapsed time is the progression of time since the beginnng of the first annoated epoch from the PSG recording. This feature might benefit from the sleep technician that arbitrarily starts annotating sleep stages from a point they deem appropriate or around the time of lights off. While this attribute might not be directly applicable to a fully automated take-home screening device, there is the near equivalent arbitrary starting point determined by the sleep participant at around the time of lights off. Participants often provide a signature to ensure synchronisation between recording devices or simply to indicate lights off. The highest ranked ECG features included RR intervals and ZCI, which are illustrative of HRV but ultimately these features do not contribute towards high REM detection sensitivity.

Alternatively, EOG features provided a boost to automated sleep staging performance (kappa score of 0.50 and mean REM F1 score of 0.52), as shown in Figure 1. These results are comparable to EOG sleep staging literature that achieved four-state kappa scores of 0.59 (Virkkala et al. 2008) and REM F1 scores of 0.64-0.80 (Agarwal et al. 2005; Virkkala et al. 2008; Yetton et al. 2016). It is clearly evident that for automated sleep staging, the ability of ECG features is superceded by EOG features. Combining EMG and EOG features, further improved automated sleep staging with a kappa of 0.57 and mean REM F1 score of 0.62. Using all three sensors (EOG, EMG, and ECG) provided an almost negligible improvement (kappa score of 0.58 and mean REM F1 score of 0.60). Essentially providing no additional merit for an additional ECG sensor, which would only add cost and hassel to a potential screening support tool. The inclusion of EEG features (Z3 combination, see Figure 1), as per our previous study (Cooray et al. 2019), provided only a marginal improvement to REM detection, but markedly improved three-stage kappa, due to improved NREM and wake classification (kappa score of 0.72 and mean REM F1 score of 0.62).

The feature ranking for REM detection, when using all signals, are detailed in Figure 2 (b). The prevalance of EOG features only confirms their correlation with REM sleep and specifically detecting rapid eye movement through permutation entropy, max peak amplitude, and coastline features. In the ranking order, these features are closely followed by elapsed time and EMG features that measure muscle atonia (amplitude percentile, entropy, and relative power). The addition of ECG features did not improve EOG, EMG or EOG and EMG combinations, indicating that these features are often redundant and perhaps only help to reduce variation (see Figure 1). While ECG features did not appear fruitful in automated sleep staging, they had some success in RBD detection.

RBD classification through ECG metrics proved effective but ultimately did not outperform EMG metrics (see Figure 3). The ECG metric ranking for RBD detection ((a)) indicates that NREM origin count and NREM irregular index as the most informative. Literature describes a loss in HRV regulation during REM (Postuma et al. 2010; Sorensen et al. 2013; Bugalho et al. 2018) and differences in atrial fibrillation (Hong et al. 2019) in individuals diagnosed with RBD. Combining EMG and ECG metics provided a marginal improvement to RBD detection, with regards to the F1 score (Figure 3). This suggests that ECG and EMG metrics are correlated, in that a reduction of HRV regulation during REM is synonymous with a loss of muscle atonia. The impact of automated sleep staging on RBD detection was also evaluated using the most successful combinations of sensors (detailed in Figure 1).

The performance of RBD detection using the best sensor combinations for both automated sleep staging and RBD classification are detailed in Figure 5. The boost in performance from using more than one sensor for sleep staging can be observed in C2E1, C2D2, and A3D2 (multiple sensors) compared to B1D1, B1E1, and B1D2 (single EOG sensor). Previously our results confirmed that C2 and A3 provide better REM detection sensitivity compared to B1 (Figure 1), thereby enabling better RBD detection through better REM classification. Interestingly, when using automated sleep staging (C2 – EOG and EMG), RBD detection using ECG metrics (C2E1) are equivalent to EMG metrics (C2D2), see Figure 5. Once again proving that there is considerable overlap in the discriminating abilities of both these metrics, where performance can be differentiated with better REM detection. In terms of achieving economy and efficiency, C2E1 used two sensors (EOG and EMG) compared to three for C2D2 and A3D2 (EOG, EMG and ECG), a potential advantage for a simple and cost-effective screening tool. Furthermore these results are comparable to results derived using EEG features illustrated by the Z3E1 combination, thereby bypassing the complications of applying EEG electrodes.

From Table 7 we can observe the RBD detection results using manually and automatically annotated sleep staging (using the minimalistic combination of a EMG and EOG sensor). While RBD detection performance dropped slightly when using automated sleep staging, it still remained high (mean accuracy, sensitivity and specificity of 0.90, 0.88, and 0.92, respectively).

Bland-Altman analysis was used to evaluate the agreement between metric quantification using manually and automatically annotated sleep staging from the C2 combination (EOG and EMG). Figure 6 (a)-(e) provides the Bland-Altman plots for the best metrics for RBD detection using the EMG sensor (atonia index, fractal exponent, and atonia index ratio) and the ECG sensor (irregular index, RR interval index ratio, and NREM origin count) as detailed in Figure 6. From these plots it is clear that automated sleep staging (derived from EOG and EMG features) sufficiently produced RBD metrics that were correlated to metrics derived from manually annotated sleep staging. However we can observe a bias introduced to RBD metrics derived from automatically annotated sleep staging, where the mean difference (solid black line in Figure 6) and its' confidence interval (grey shaded area) does not include the line of equality (zero axis, dotted line through the origin). This is the direct result from automated sleep stage misclassification. When REM is misclassifed for NREM and wake, the impact on the calculation of RBD metrics provides a bias towards values associated with NREM and wake sleep. Because automated sleep staging was poorest on RBD participants (red circles), we observed a greater variation in the y-axis (difference), especially as the x-axis values (mean value) moved towards values associated with RBD (red circles). Note that the limits of agreement are calculated by assuming that the difference between the metrics calculated from manually and automatically annotated sleep stages are normally distributed, however this isn't strictly necessary (Bland and Altman 1999). From these plots we can observe the resilience and fidelity of these metrics using automated sleep staging from the most successful combination of minimal PSG sensors (EOG and EMG).

These results validate a cost-effective and readily accessible take-home sleep device for RBD screening purposes using only an EOG and EMG sensor. However, often epidemiological studies incur much greater costs for logistical reasons and include additional sensors to optimise their collection of data and therefore include ECG and EEG sensors. Often wearable devices and general sleep recordings are plagued with inconsistent and intermittent signals, given the unpredictability and difficulty of self-administered devices outside of a clinical setting. This study overcomes these issues and further supports retrospective and prospective studies by demonstrating the redundancy and ability of these sensors to function interchangeably for both automated sleep staging and RBD detection.

## 6. Limitations & Future Direction

While this study included a high number of participants, its application in a clinical setting are limited because it simply focused on HC and RBD cohorts. A clinical application would demand greater resilience to a myriad of sleeping disorders and other population variations that could easily share confounding RBD attributes. These include periodic limb movement of sleep, OSA, parasomnias, and severe insomnia (very fragmented sleep), which should also be included and evaluated. While ECG metrics did not prove as effective in RBD detection as EMG metrics, there is still no clear indication of which characteristic manifests earlier in RBD participants, loss of REM atonia or loss of HRV regulation during sleep. The loss of HRV regulation might even provide an early bio-marker for RBD, and could potentially offer insight into the PD conversion of RBD participants, although literature exists that suggests there is currently no correlation for the latter (Postuma et al. 2010). This study also confirmed that successful REM detection is paramount to effective RBD detection and future work to improve automated REM detection would only prove beneficial. Advances in deep learning techniques are achieving promising results and may prove useful in this context (Supratak et al. 2017; Phan et al. 2019). Furthermore applying these tools on wearable data would be the next logical step forward towards validating a fully automated RBD detection pipeline.

## 7. Conclusion

This study proved the feasibility of a fully automated pipeline for RBD detection using an EOG and EMG sensor. This study achieved automated sleep staging comparable to manual annotation, which translated to a high performance in RBD detection. Once more this study verified RBD detection through ECG based metrics (NREM irregular index, RR interval index ratio and REM RR interval standard deviation) were effective but did not out-perform EMG based metrics. Furthermore their use in automated sleep staging provided very little benefit and may not be worth the additional sensor for a minimalistic and economical RBD screening tool.

# Table & Figures

| Database | Cohort | Age | #Subjects | #Female | #Male |
|---|---|---|---|---|---|
| CAP | RBD | 70.2±5.3 | 14 | 2 | 12 |
| JR | RBD | 64.3±8.0 | 36 | 2 | 34 |
| **Combined (CAP/JR)** | **RBD** | **68.0±7.8** | **50** | **4** | **46** |
| **MASS** | **Elderly-HC** | 63.7±5.2 | 50 | 17 | 33 |

*Table 1: Datasets used in the study. Recordings from individuals with RBD are collected from the JR and CAP datasets and have been combined (CAP/JR). Note the male predominance within the RBD dataset. These datasets are analysed separately for sleep stage classification and combined for abnormal EMG calculation and RBD detection.*

| Channel | Category | Name | Description | Reference |
|---|---|---|---|---|
| **ECG** | Non-linear | IrrIndex | Irregular Index, a feature derived from Lorenz plots from RR intervals. Originally termed atrial fibrillation evidence. | (Sarkar et al. 2008) |
| **ECG** | Time | RR Interval | The mean and median RR interval. Calculated also for normalised RR intervals | (Andreotti et al. 2018) |
| **ECG** | Time | SDNN | The standard deviation of the standard deviation of the NN interval. Also a normalised value is calculated. | (Andreotti et al. 2018) |
| **ECG** | Time | RMSSD | Square root of mean of squares of difference between adjacent NN interval | (Andreotti et al. 2018) |
| **ECG** | Time | SDSD | Standard deviation of differences between adjacent NN intervals | (Andreotti et al. 2018) |
| **ECG** | Time | NN50 | Number of pairs of adjacent NN intervals differing by more than 50ms | (Andreotti et al. 2018) |
| **ECG** | Time | pNN50 | Percentage NN50 (NN50/total number of NN intervals) | (Andreotti et al. 2018) |
| **ECG** | Frequency | LFpeak | Frequency of maximum peak in low frequency range. Also calculated with normalised value. | (Andreotti et al. 2018) |
| **ECG** | Frequency | HFpeak | Frequency of maximum peak in high frequency range. Also calculated with normalised value. | (Andreotti et al. 2018) |
| **ECG** | Frequency | Totalpower | Also calculated with normalised value. | (Andreotti et al. 2018) |
| **ECG** | Frequency | LFpower | Also calculated with normalised value. | (Andreotti et al. 2018) |
| **ECG** | Frequency | HFpower | Also calculated with normalised value. | (Andreotti et al. 2018) |
| **ECG** | Frequency | nLF | Percentage of low and high frequency power in the low frequency range | (Andreotti et al. 2018) |
| **ECG** | Frequency | nHF |  | (Andreotti et al. 2018) |
| **ECG** | Frequency | LFHF | Ratio of low to high frequency power. | (Andreotti et al. 2018) |
| **ECG** |  | PoincareSD1 | Standard deviation in y=-x direction of Poincare plot. Also calculated with normalised value. | (Andreotti et al. 2018) |
| **ECG** |  | PoincareSD2 | Standard deviation in y=x direction of Poincare plot. Also calculated with normalised value. | (Andreotti et al. 2018) |
| **ECG** | Non-linear | Sample Entropy | Sample entropy. Also calculated with normalised value. | (Andreotti et al. 2018) |
| **ECG** | Non-linear | Approximate Entropy | Approximate entropy. Also calculated with normalised value. | (Andreotti et al. 2018) |
| **ECG** | Time | RR | Recurrence rate. | (Andreotti et al. 2018) |
| **ECG** |  | DET | Determinism | (Andreotti et al. 2018) |
| **ECG** | No-linear | ENTR | Shannon entropy. | (Andreotti et al. 2018) |
| **ECG** |  | L | Average diagonal line length. | (Andreotti et al. 2018) |
| **ECG** | Time | TKEO | Mean Teager-Kaiser energy operator. Also calculated with normalised value. | (Andreotti et al. 2018) |
| **ECG** |  | DAFa2 | Detrended fluctuation analysis exponent. | (Andreotti et al. 2018) |
| **ECG** |  | LZ | Lempel Ziv complexity. | (Andreotti et al. 2018) |
| **ECG** |  | BD | Mutual information | (Andreotti et al. 2018) |
| **ECG** |  | PD | Mutual information | (Andreotti et al. 2018) |
| **ECG** |  | BDa | Auto-correlation | (Andreotti et al. 2018) |
| **ECG** |  | PDa | Auto-correlation | (Andreotti et al. 2018) |
| **ECG** | Time | ZCI | Zero crossing interval. | (Andreotti et al. 2018) |
| **ECG** | Time | mZCI | Mean zero crossing interval. | (Andreotti et al. 2018) |
| **ECG** | Time | nsZCI | Normalised zero crossing interval. | (Andreotti et al. 2018) |
| **ECG** | Time | AmpVarsqi | Variation of amplitude around QRS complex. | (Andreotti et al. 2018) |
| **ECG** | Time | Ampstdsqi | Standard deviation of amplitude around QRS complex. | (Andreotti et al. 2018) |
| **ECG** | Time | AmpMean | Mean amplitude around QRS complex. | (Andreotti et al. 2018) |
| **ECG** | Time | Tachy | Tachycardia ( > 100 beats per minute (bpm) in adults) | (Andreotti et al. 2018) |
| **ECG** | Time | Brady | Bradycardia ( < 60 bpm in adults) | (Andreotti et al. 2018) |
| **EOG** | Time | Percent Differential | This feature is calculated by taking the difference between the 75th and the 25th percentile. The 75th and 25th percentile is defined by the amplitude, below which represents 75% and 25% of the random value, respectively. | (Lajnef et al. 2015) |

| Channel | Category | Name | Description | Reference |
|---------|----------|------|-------------|-----------|
| EOG | Time | Mean, Minimum, & Maximum Coastline | Calculated by summating the absolute derivatives of the signal. The coastline is determined for each mini-epoch, where the mean, minimum and maximum can be calculated for a given 30s epoch. | (Yetton et al. 2016) |
| EOG | Time | Autocorrelation | The peak of the autocorrelation sequence of every mini-epoch was calculated. The autocorrelation gives an indication of a pattern within a given mini-epoch, enhancing REM spikes within the EOG signal. | (Yetton et al. 2016) |
| EMG/EOG | Time | Variance | The variance of each mini-epoch. | |
| EOG | Time | Kurtosis | The measure of kurtosis for each epoch. | |
| EOG | Time | Skewness | Measure of skewness for each epoch. | |
| EOG | Non-linear | Teager-Kaiser Energy Operator | The mean of the Teager-Kaiser Energy Operator (TKEO) was calculated. | (Lajnef et al. 2015) |
| EOG | Time | Max Peak | The maximum absolute peak of each mini-epoch was detected as a feature (positive or negative). | (Yetton et al. 2016) |
| EOG | Time | Second Max Peak | The second largest maximum absolute peak of each mini-epoch was detected as a feature (positive or negative). | (Yetton et al. 2016) |
| EOG | Time | Peak Prominence | This is a measure of how much peaks stand out compared to other surrounding peaks. | (Yetton et al. 2016) |
| EOG | Time | Peak Width | The width of a peak is the distance between two points, left and right of the peak, which intercept a vertical distance from the peak of half the peak prominence. | (Yetton et al. 2016) |
| EOG | Time | Rise | The slope on the rise to the peak. | (Yetton et al. 2016) |
| EOG | Time | Fall | The slope on the fall from the peak. | (Yetton et al. 2016) |
| EOG | Time | Mean, maximum, and variance of Differential | The average differential value is calculated for every mini-epoch, where the average and the maximum value across the 30s epoch is determined. | (this work) |
| EOG | Frequency | Power Ratio (0.5Hz/4Hz) | The power ratio between the 0-4Hz band and the entire spectrum was calculated. | (Susmáková and Krakovská 2008) |
| EOG | Wavelet | Discrete Wavelet Transform (DWT) - Haar/DB2 | The DWT was applied using 4 levels of decomposition and used the maximum amplitude of the -4 level inverse DWT as the feature. The larger the amplitude of the inverse DWT signal the more the signal can be composed of wavelets and also more likely that the window contains REM | (Yetton et al. 2016) |
| EOG | Non-linear | Permutation Entropy | Permutation entropy is a nonlinear measure that characterises the complexity of a time series. For every mini-epoch the permutation entropy was calculated to the 10th order. | (Lajnef et al. 2015) |
| EMG | Time | Atonia Index | The distribution of amplitude values (averaged and corrected) for a given 30s epoch are used to determine the percentage of values ≤ 1µV (excluding values > 1µV and < 2µV). | (Ferri et al. 2010) |
| EMG | Time | Energy | For every epoch the mean rectified amplitude is calculated. | (Hsu et al. 2013) |
| EMG | Time | 75th Percentile | The 75th percentile, detailing the value which 75% of the variable is below. | (Charbonnier et al. 2011) |
| EMG | Time | Entropy | The variability of the signal is calculated as follows, where $P_i$ is the histogram count of values using 256 bins: | (Charbonnier et al. 2011) |
| EMG | Time | Motor Activity | An algorithm that derives a signal to measure motor activity and a threshold (baseline) to determine when there is movement. The threshold is then used to determine the duration of motor activity within each epoch. | (Frandsen et al. 2015) |
| EMG | Frequency | Fractal Exponent | The negative slope of the spectral density using a logarithmic on both the frequency and power. $P(f) \sim f^{-\alpha}$ | (Susmáková and Krakovská 2008) |
| EMG | Frequency | Absolute Gamma Power | The average power in the gamma frequency range (30-100Hz). | (Susmáková and Krakovská 2008) |
| EMG | Frequency | Relative Power | The ratio of the average frequency power between the high frequency range (12.5-21Hz) and the total frequency band (8-32Hz). | (Charbonnier et al. 2011) |
| EMG | Frequency | Spectral Edge Frequency | The highest frequency at which 95% of the total signal power is located. | (Charbonnier et al. 2011) |
| EMG | Time | Standard Deviation | The standard deviation of the amplitude for each epoch. | |
| NA | Time | Hours Recorded | The nature of sleep architecture means the progress of time provides details on the likelihood of having REM, therefore the number of hours into PSG recordings is also used as a feature. | (this work) |
| NA | Time | Hours From End | Time in hours from the end of the recording. This features provides information on likelihood of wake and REM. | (this work) |

*Table 2: Features extracted from ECG, EOG, and EMG signals.*

| Channel | Category | Name | Description | Reference |
|---|---|---|---|---|
| **ECG** | Time | RR_REM_Std | The mean RR interval standard deviation during all REM epochs. | (Andreotti et al. 2018) |
| **ECG** | Frequency | LFpeak_NREM | The mean RR interval low frequency peak during NREM epochs. | (Andreotti et al. 2018) |
| **ECG** | Frequency | LFpeak_REM | The mean RR interval low frequency peak during REM epochs. | (Andreotti et al. 2018) |
| **ECG** | Frequency | HFpeak_NREM | The mean RR interval high frequency peak during NREM epochs. | (Andreotti et al. 2018) |
| **ECG** | Frequency | HFpeak_REM | The RR interval high frequency peak during REM epochs. | (Andreotti et al. 2018) |
| **ECG** | Time | RR_Index_REMNREM | The mean RR interval ratio between REM and NREM epochs. | This work. |
| **NA** | Time | Ratio_NREM | Ratio of NREM sleep compared to all other stages. | This work |
| **ECG** | Frequency | LFHF_Index_REMNREM | The mean RR interval low frequency to high frequency ratio between NREM and REM epochs. | This work |
| **ECG** | Non-linear | IrrIndex_NREM | The mean irregular index. Originally termed atrial fibrillation evidence. | (Sarkar et al. 2008) |
| **ECG** | Non-linear | IrrIndex_REM | The mean irregular index calculated for each REM epoch. Originally termed atrial fibrillation evidence. | (Sarkar et al. 2008) |
| **ECG** | Non-linear | OriginCount_NREM | A parameter calculated for the irregular index that measured the number of regular heart beat intervals for each 30s epoch and then averaged using a sliding 150s window. This metric calculates the mean origin count for all NREM epochs. | (Sarkar et al. 2008) |
| **ECG** | Non-linear | OriginCount_REM | Similar to Origin_Count_NREM, where this metric calculates the mean origin count for all REM epochs. | (Sarkar et al. 2008) |
| **ECG** | Non-linear | SampEn_REM | The mean sample entropy during REM. | (Andreotti et al. 2018) |

*Table 3: ECG metrics for RBD detection, which aim to capture changes in HRV during sleep.*

| **Sleep Staging Signals** | **#Features** | **#Sensors** | **ID** |
|---|---|---|---|
| ECG | 75 | 1 | A1 |
| EOG | 25 | 1 | B1 |
| EMG | 17 | 1 | C1 |
| ECG+EOG | 98 | 2 | A2 |
| ECG+EMG | 90 | 2 | B2 |
| EOG+EMG | 40 | 2 | C2 |
| ECG+EOG+EMG | 113 | 3 | A3 |
| EEG+EOG+EMG | 128 | 3 | Z3 |

*Table 4: Combination of austere PSG sensors to be used and analysed for automated sleep staging (three state). The Z3 combination is based on our previous study and used as a comparison* (Cooray et al. 2019).

| **RBD Detection Signals** | **#Features** | **Total Sensors** | **ID** |
|---|---|---|---|
| ECG | 11 | 1 | D1 |
| EMG | 7 | 1 | E1 |
| ECG+EMG | 18 | 2 | D2 |

*Table 5: These are the austere PSG sensors to be analysed for RBD detection in coordination with automated sleep staging.*

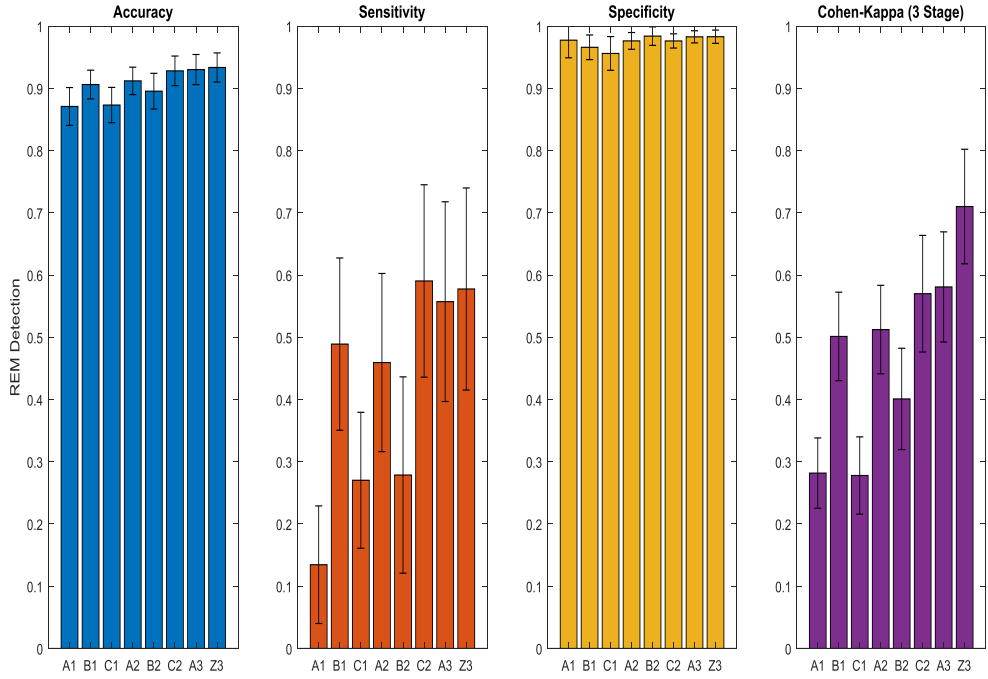

*Figure 1: The depiction of REM detection performance for various PSG signal combinations. For a single sensor (A1, B1, and C1) the best sensitivity and F1 score is given by the EOG signal (B1). For two sensors, (A2, B2, C2), the best performance is given by EOG and EMG (C2). This exceeds the performance of a single EOG sensor (B1), which is illustrated by the increase in sensitivity, while maintaining an equal specificity.*

|             | W           | NREM        | REM         |
|-------------|-------------|-------------|-------------|
| **Accuracy**    | 0.79±0.09   | 0.72±0.08   | 0.87±0.06   |
| **Sensitivity** | 0.42±0.17   | 0.90±0.09   | 0.13±0.19   |
| **Specificity** | 0.91±0.08   | 0.37±0.16   | 0.98±0.06   |
| **Precision**   | 0.59±0.19   | 0.72±0.10   | 0.38±0.35   |
| **F1**          | 0.45±0.12   | 0.80±0.08   | 0.16±0.19   |
| **Kappa**       | 0.34±0.13   | 0.30±0.13   | 0.13±0.18   |

(a)

|             | W           | NREM        | REM         |
|-------------|-------------|-------------|-------------|
| **Accuracy**    | 0.85±0.10   | 0.81±0.098  | 0.93±0.047  |
| **Sensitivity** | 0.64±0.21   | 0.90±0.12   | 0.60±0.31   |
| **Specificity** | 0.92±0.11   | 0.66±0.18   | 0.98±0.022  |
| **Precision**   | 0.73±0.20   | 0.83±0.096  | 0.76±0.22   |
| **F1**          | 0.64±0.17   | 0.85±0.097  | 0.62±0.27   |
| **Kappa**       | 0.56±0.19   | 0.57±0.18   | 0.59±0.27   |

(b)

*Table 6: Performance of automatic sleep stage classification using (a) ECG features and (b) combined EOG and EMG features. For (a) performance is relatively low compared to literature on ECG sleep staging, but interestingly this is overwhelmingly due to performance on RBD participants, where sensitivity is many times smaller than HCs. The performance of (b) is substantially better than using an ECG feature and a single EOG feature.*

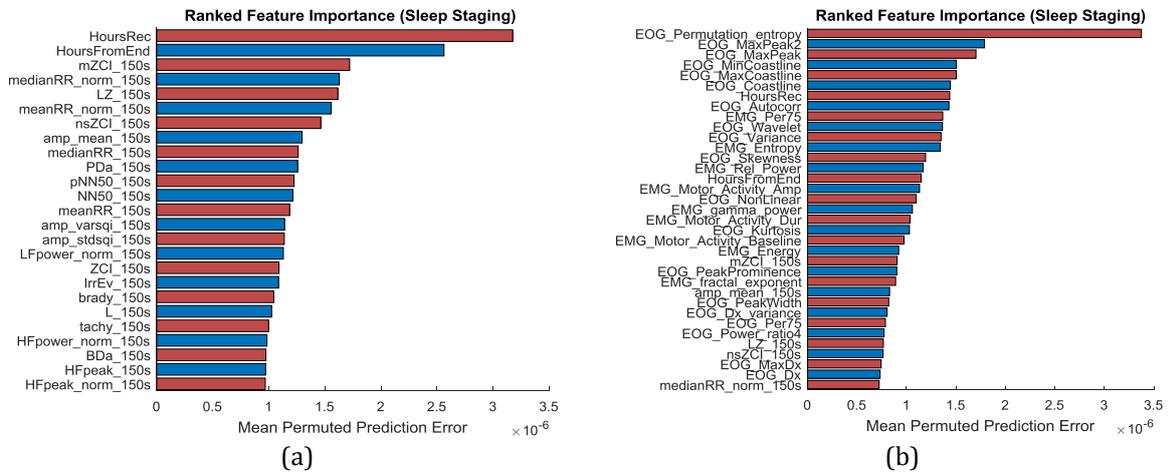

*Figure 2: The order of feature importance for automated sleep staging (REM detection) using (a) ECG features compared with (b) ECG, EOG, and EMG features. The elapsed recording time represents the top two features for (a) followed by ZCI and RR interval based features. Also highly represented are the ECG amplitude based features such as amplitude mean and, variance and stdsqi. For (b) the EOG features prove most important for REM detection, specifically permutation entropy, max peak*

*and coastline features. Followed by elapsed recording time and EMG features, such as the 75th percentile, entropy, relative power and motor activity. These additional features from another signal provide the boost in REM detection performance shown in Figure 1.*

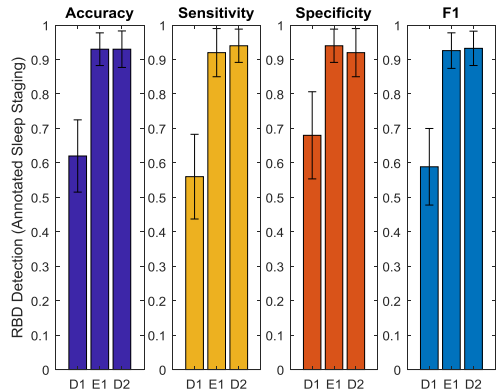

*Figure 3: Performance of RBD detection using ECG (D1), EMG (E1), and both ECG and EMG signals (D2) with manually annotated sleep stages. With annotations ECG features (D1) prove effective at RBD identification, but are clearly outperformed by an EMG sensors (E1). Combining these sensors (D2) only improves performance marginally with respect to the F1 score.*

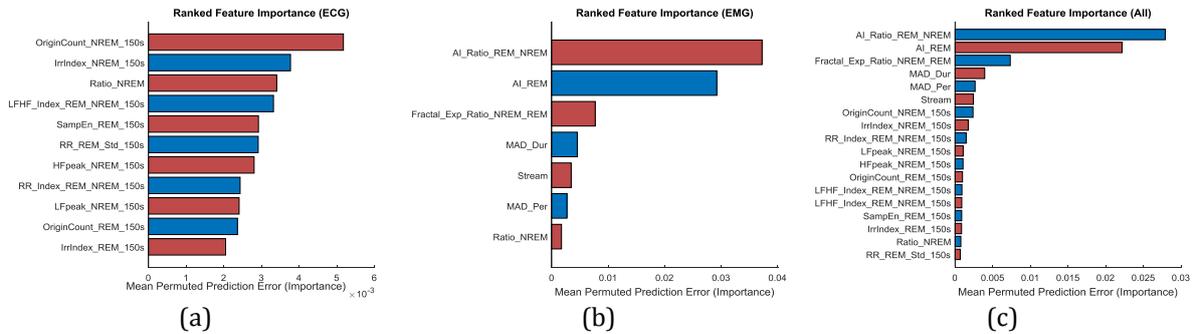

*Figure 4: Features importance for RBD detection using (a) ECG, (b) EMG and (c) both EMG and ECG metrics. From the ECG features (a), irregular evidence during NREM proves the most effective for RBD detection, followed by the LFHF index ratio between REM and NREM.*

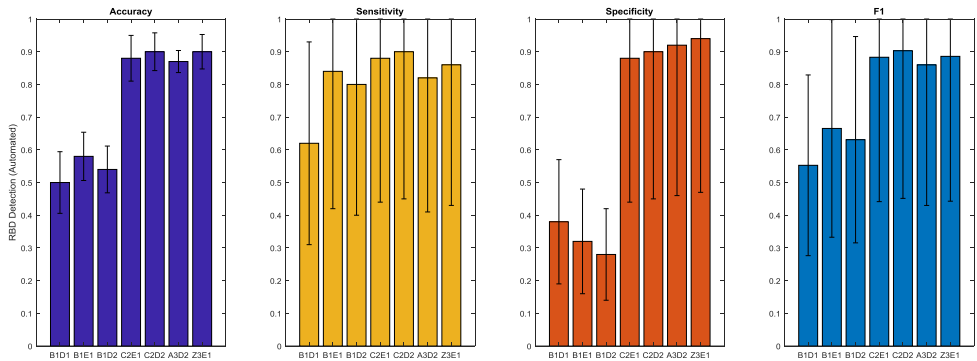

*Figure 5: RBD Detection performance when using automated sleep staging. For this figure B1, C2 and A3 represent automated sleep staging using only EOG, both EOG and EMG, or ECG, EOG and EMG features, respectively. D1, E1 and D2 represent metrics derived from EMG, ECG, and both EMG and ECG sensors for RBD detection, used in combination with automated sleep staging.*

| RBD Detection | RBD Detection using Automatic Sleep Staging | | | |
|---|---|---|---|---|
| | Accuracy | Sensitivity | Specificity | F1 |
| **Atonia Index** | 0.83±0.12 | 0.68±0.22 | 0.98±0.06 | 0.78±0.18 |
| **RF (EMG Features)** | 0.90±0.11 | 0.88±0.13 | 0.92±0.098 | 0.90±0.12 |
| **RF (ECG Features)** | 0.65±0.22 | 0.70±0.24 | 0.60±0.24 | 0.66±0.21 |
| **RF (EMG+ECG Features)** | 0.92±0.098 | 0.88±0.13 | 0.96±0.08 | 0.91±0.11 |

*Table 7: RBD Detection results using automatically annotated sleep stages. Automated sleep stages achieved using EOG and EMG features.*

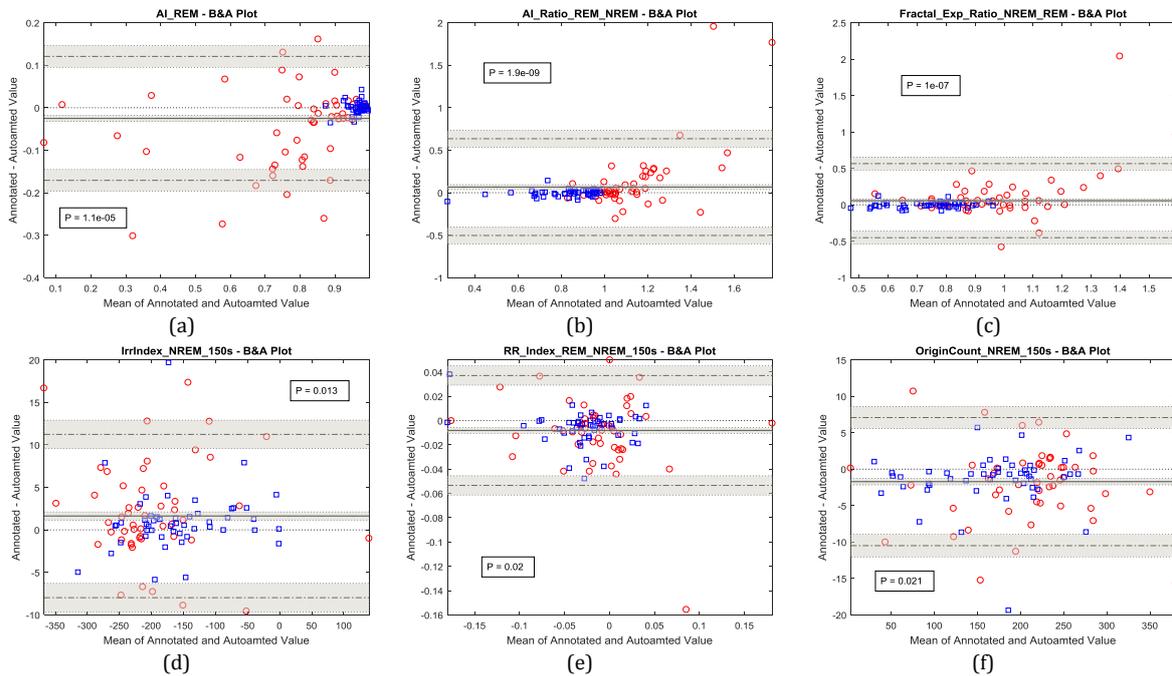

*Figure 6: Bland and Altman plots of important RBD metrics comparing scores derived from manually and automatically annotated sleep stages. These include the (a) atonia index during REM, (b) atonia index ratio between REM and NREM, (c) fractal exponent ratio between REM and NREM, (d) irregular index during NREM, (e) RR interval index between REM and NREM, and (f) origin count during NREM. In all cases we can observe that metrics are within the limits of agreement. The P-value details the Kolmogorov-Smirnov test to determine if the difference between metrics calculated from manual and automatic sleep staging is from a normal distribution (rejected when < 0.05). For metrics shown above, the line of equality (the zero dotted line) doesn't fall within the limits of confidence around the mean difference. Therefore when using automated sleep staging the calculated metrics for RBD detection slightly shifts towards values associated with RBD or HC participants. This is unsurprising given misclassification from automated sleep staging will have an impact on the calculation of these metrics.*